\documentclass[prl,aps,twocolumn,showpacs,floatfix,longbibliography]{revtex4-2}
\usepackage{amsmath}
\usepackage{amsfonts}
\usepackage{amssymb}
\usepackage{hyperref}
\usepackage{graphicx}
\usepackage{dcolumn}
\usepackage{bm,amsmath,verbatim}
\usepackage{mathrsfs}
\usepackage{color}

\usepackage[T1]{fontenc}
\usepackage[latin9]{inputenc}
\usepackage{booktabs}

\allowdisplaybreaks[2]

\hypersetup{colorlinks,
	linkcolor=blue,          citecolor=blue,        filecolor=blue,      urlcolor=blue           }

\begin{document}

\title{Fractional Quantum Hall Effect Based on Weyl Orbits}

\author{Jiong-Hao Wang$^{1}$}
\author{Yan-Bin Yang$^{1}$}
\author{Yong Xu$^{1,2}$}
\email{yongxuphy@tsinghua.edu.cn}
\affiliation{$^{1}$Center for Quantum Information, IIIS, Tsinghua University, Beijing 100084, People's Republic of China}
\affiliation{$^{2}$Hefei National Laboratory, Hefei 230088, People's Republic of China}
	
\begin{abstract}
The fractional quantum Hall effect is a well-known demonstration of strongly correlated topological phases in two dimensions. 
However, the extension of this phenomenon into a three-dimensional context has yet to be achieved.
Recently, the three-dimensional integer quantum Hall effect based on Weyl orbits has been experimentally 
observed in a topological semimetallic material under a magnetic field.
This motivates us to ask whether the Weyl orbits can give rise to the fractional quantum Hall effect when their Landau level is 
partially filled in the presence of interactions. 
Here we theoretically demonstrate that the fractional quantum Hall states based on Weyl orbits can emerge in a Weyl semimetal
when a Landau level is one-third filled.
Using concrete models for Weyl semimetals in magnetic fields,
we project the Coulomb interaction onto a single Landau level from the Weyl orbit and 
find that the ground state of the many-body Hamiltoian is triply degenerate. We further show that 
the ground states exhibit the many-body Chern number of $1/3$ and
the uniform occupation of electrons in both momentum and real space, implying that they are the fractional quantum Hall states.
In contrast to the two-dimensional case, the states are spatially localized on two surfaces hosting Fermi arcs. 
Additionally, our findings suggest that the excitation properties of these states resemble those of the Laughlin state in two dimensions, 
as inferred from the particle entanglement spectrum of the ground states.
\end{abstract}
\maketitle

\section{Introduction}

The fractional quantum Hall (FQH) effect~\cite{jain1992AdvPhys,stormer1999RMP,
	stormer1999nobel,murthy2003RMP,hansson2017RMP}, originating from the interplay between 
topology and strong correlations, has been one of the central topics in condensed matter physics
since its discovery~\cite{tsui1982PRL}.
The FQH states possess exotic properties including fractionally quantized Hall 
conductance~\cite{tsui1982PRL,niu1985PRB} as well
as anyonic excitations with fractional charges~\cite{laughlin1983PRL,arovas1984PRL,halperin1984PRL}.
They serve as a paradigmatic example of topological order~\cite{wen1990PRB,wen1995AdvPhys,wen2017RMP} with long-range 
entanglement~\cite{kitaev2006PRL,levin2006PRL} 
and have potential applications in topological quantum computation~\cite{read1991NPB,read1999PRB,nayak2008RMP}.
Although the study of FQH states has lasted for decades, it still awards us with new surprise,
for example, the geometric degree of freedom~\cite{haldane2011PRL} that gives rise to 
chiral gravitons~\cite{yang2012PRL,golkar2016JHEP,liou2019PRL,liang2024Nature}. 
Recently, the FQH effect has been realized in various quantum simulators with high 
controllability, including photons~\cite{clark2020Nature,wang2024Science} and cold atoms~\cite{leonard2023Nature}.
Its counterpart under no magnetic field, 
called fractional quantum anomalous Hall state or fractional Chern 
insulator~\cite{neupert2011PRL,sheng2011NC,wang2011PRL,regnault2011PRX,tang2011PRL,sun2011PRL}, 
has been observed experimentally in moir\'{e} superlattice structures including twisted 
bilayer transition metal dichalcogenidematerial MoTe$_2$~\cite{cai2023Nature,zeng2023Nature,park2023Nature,xu2023PRX} 
and rhombohedral pentalayer graphene-hBN heterostructure~\cite{lu2024Nature}.
The progress provides new platforms for the study of the FQH states and 
gives chance to uncover more hidden treasure in this fascinating quantum matter. 
Despite the substantial advancements in experiments and theoretical understanding of the FQH states,
the existing research focuses on two-dimensional (2D) systems, and the FQH physics in three dimensions (3D) is still
elusive because the energy dispersion due to the additional dimension melts the separation of Landau
levels, disfavoring the occurrence of the FQH effect.   

Weyl semimetals~\cite{jia2016NM,burkov2016NM,armitage2018RMP,xu2019FP,lv2021RMP} 
host low-energy excitations with linear dispersions near the gapless Weyl points which come 
in pairs with positive and negative topological charges.
The topological surface states named Fermi arcs connect Weyl points with opposite chirality 
and result in anomalous Hall conductivity~\cite{yang2011PRB,burkov2011PRL}.
Remarkably, under the magnetic field, the Fermi arcs on opposite surfaces form the Weyl orbit~\cite{potter2014NC}
in the assistance of bulk Weyl points, leading to Landau levels supporting the 3D quantum Hall effect~\cite{wang2017PRL,zhang2017NC,uchida2017NC,schumann2018PRL,zhang2019Nature,
	lu2019NSR,li2020PRL,nishihaya2021NC,chen2021PRL,zhang2021NRP,li2021npjQ,chang2022CPB,xiong2022PRB,zhangxr2023PRB,wang2023arXiv}.
It is thus natural to ask whether the Weyl orbit can bring about the 3D FQH effect
when the Landau level is partially filled with appropriate filling factor and the interaction is turned on.
The answer is unclear because the Landau level emerging from the Weyl orbit is very different from 
the lowest Landau level in 2D (see Appendix B).

In this work, we theoretically demonstrate that the FQH effect based on Weyl orbits can arise in a Weyl semimetal under the magnetic field.
By projecting the Coulomb interaction onto a Landau level
originating from Weyl orbits, we obtain the energy spectrum of the many-body Hamiltonian using exact diagonalization
for a one-third filled case.
We find that the triply degenerate ground states occur in the 
specific total momentum sectors predicted by the generalized Pauli principle, 
which are identified as the FQH states justified by the many-body Chern number.
The possibility of the charge density wave (CDW) is excluded because the fluctuation of the occupation number 
on each single-particle state is small and the real-space electron density is quite uniform in the plane
parallel to the top and bottom surfaces, characteristic of a FQH fluid.
The electron fluid concentrates on the top and bottom surfaces, which is a distinctive feature of the 
FQH effect based on Weyl orbits, indicating its 3D nature.
Despite the significant difference in the spatial character 
from the conventional FQH effect in a 2D electron gas,
the excitation properties of the Weyl-orbit-based FQH states resemble those of the Laughlin state,
manifesting in the particle entanglement spectrum.
Finally, as real Weyl materials usually host multiple pairs of Weyl points, 
we study a time-reversal symmetric Weyl semimetal model with two pairs of Weyl points
under a magnetic field 
and find that the FQH effect can also occur in this system based on Weyl orbits.  

\section{Model}

\begin{figure}[t]
	\includegraphics[width=3.4in]{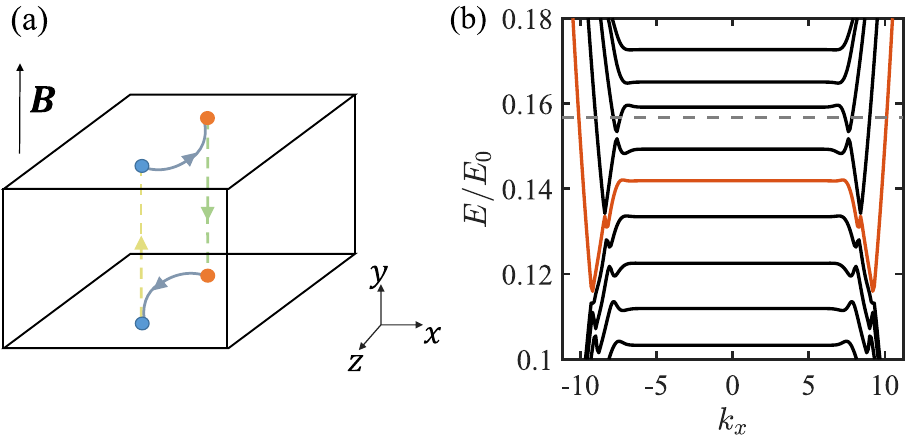}
	\caption{
		(a) Schematic illustration of Fermi arcs and Weyl orbits formed under the magnetic field ${\bm B}$ 
		for the Hamiltonian (\ref{Ham1}).
		(b) The Landau levels of the Hamiltonian (\ref{Ham1}) 
		near the energy of the Weyl points
		(the dashed grey line).
		The Landau level onto which the Coulomb interaction is projected is colored red.
		Here we take OBCs along $y$ and $z$ and PBCs along $x$.
		The system size used for the calculation is $\widetilde{L}_y=L_y/(\sqrt{2\pi}l_B)=4$
		and $\widetilde{L}_z=L_z/(\sqrt{2\pi}l_B)=8$ with $L_y$ and $L_z$ the length of the system along 
		$y$ and $z$, respectively.
	}
	\label{Fig1}
\end{figure}

To investigate the FQH effect based on Weyl orbits, 
we consider the following single-particle Hamiltonian of a Weyl semimetal in momentum space~\cite{wang2017PRL}
\begin{eqnarray}\label{Ham1}
H(\bm{k})&=&M[k_w^2-(k_x^2+k_y^2+k_z^2)]\sigma_z+A(k_x \sigma_x +k_y\sigma_y) \nonumber \\
&&+[D_1k_y^2+D_2(k_x^2+k_z^2)]\sigma_0,
\end{eqnarray} 
where $\bm{k}=(k_x,k_y,k_z)$ is the wave vector, $\sigma_\nu$ with $\nu=0,x,y,z$ are identity and Pauli matrices,
and $k_w$, $M$, $A$, $D_1$ and $D_2$ are real system parameters.
The model hosts a pair of Weyl points at ${\bm{k}}=(0,0,\pm k_w)$ in the bulk with the energy $E_W=D_2k_w^2$.
Under open boundary conditions (OBCs) along $y$,
the projections of the bulk Weyl points onto the top and bottom surfaces are connected by bent Fermi arcs
as shown schematically in Fig.~\ref{Fig1}(a).
The Fermi arcs lead to the Weyl orbit [a closed loop shown in Fig.~\ref{Fig1}(a)] in the presence 
of magnetic fields $\bm{B}=(0,B,0)$ along $y$~\cite{potter2014NC,wang2017PRL,li2020PRL}.
The Weyl orbit gives rise to Landau levels around the energy of the Weyl points, as shown in Fig.~\ref{Fig1}(b) 
(see Appendix A for the basis used to calculate the Landau levels).
Here we set the parameters as $\widetilde{M}=M/l_B^2=0.04E_0,\widetilde{D}_1=D_1/l_B^2=0.01E_0,
\widetilde{D}_2=D_2/l_B^2=0.02E_0,\widetilde{A}=A/l_B=0.04E_0$ and $\widetilde{k}_w=k_wl_B=2.8$,
where $l_B=\sqrt{\hbar/(eB)}$ is the magnetic length with $\hbar$ being the reduced Planck constant 
and $e$ being the elementary charge, and $E_0$ is the unit of energy.
The above set of parameters are used throughout the paper.

To study the FQH effect, we include the following Coulomb interaction between electrons
under periodic boundary conditions (PBCs) along $x$ and $z$ and OBCs along $y$,
 \begin{equation}\label{Coulomb}
 V(\bm{r})=\sum_{t,s=-\infty}^{+\infty}\frac{e^2}{4\pi\epsilon}\frac{1}{\bm{r}+tL_x\bm{e_x}+
 	sL_z\bm{e_z}},	
 \end{equation} 
where $t$ and $s$ take integer values, $\epsilon$ is the dielectric constant,
$\bm{e}_x$ ($\bm{e}_z$) are the unit vectors along $x$ ($z$), and 
$L_x$ ($L_z$)
is the length of the system along $x$ ($z$).
Since the sum of the Chern number of all the bands below the Landau level highlighted by the red line in Fig.~\ref{Fig1}(b) 
vanishes, we consider the case that this Landau level is one-third filled.
Similar to the FQH effect in 2D or the fractional Chern insulator, in numerical calculations, 
we project the Coulomb interaction onto the Landau level
and obtain the following many-body Hamiltonian 
 \begin{equation}\label{Ham2}
 	\hat{H}_I=\sum_{j_1,j_2,j_3,j_4=0}^{m_L-1}B_{j_1j_2j_3j_4}\hat{a}^\dagger_{j_1}\hat{a}^\dagger_{j_2}
 	\hat{a}_{j_3}\hat{a}_{j_4},	
 \end{equation} 
where $\hat{a}^\dagger_{j_\mu}$ ($\hat{a}_{j_\mu}$) is the creation (annihilation) operator 
corresponding to the single-particle state with momentum $k_x=2\pi j_\mu/L_x$ in the considered
Landau level. 
Here, the momentum quantum number $j_\mu$ with $\mu=1,2,3,4$ takes integer values from $0$ to $m_L-1$ where 
$m_L=L_xL_z/(2\pi l_B^2)=\widetilde{L_x}\widetilde{L_z}$ is the degeneracy of the Landau level with
$\widetilde{L_x}=L_x/(\sqrt{2\pi}l_B)$.   
The explicit form and detailed derivation of the coefficients $B_{j_1j_2j_3j_4}$ can be found in Appendix C.
The single-band projection method used here is commonly used in the study of the FQH effect and fractional
Chern insulators, see, e.g., ~\cite{sheng2011NC,yoshioka1983PRL,wang2024PRL}.
Note that we neglect the single-particle energy including the kinetic energy and the potential energy
due to the interaction between an electron and its image under PBCs~\cite{yoshioka1983PRL} 
in the Hamiltonian 
(\ref{Ham2}),
because it only contributes a constant and does not alter the physics when the particle number is 
conserved. 

\section{Many-body energy spectrum}

To demonstrate the existence of the FQH effect based on Weyl 
orbits, we diagonalize the many-body Hamiltonian (\ref{Ham2}) for the electron number $n_e=8,10$ 
in a system with the degeneracy of the Landau level $m_L=24,30$, respectively.
Since the Hamiltonian (\ref{Ham2}) is translationally invariant, 
$K_x$ (the sum of the momentum quantum number $j$ of all occupied single-particle states modulo $m_L$)
is a good quantum number.
In the calculations,
we fix the thickness of the system to be $\widetilde{L}_y=4$
and take the aspect ratio $r_a=L_x/L_z$ to be $r_a=1,0.8$ for $n_e=8,10$,
respectively.

\begin{figure}[t]
 	\includegraphics[width=3.4in]{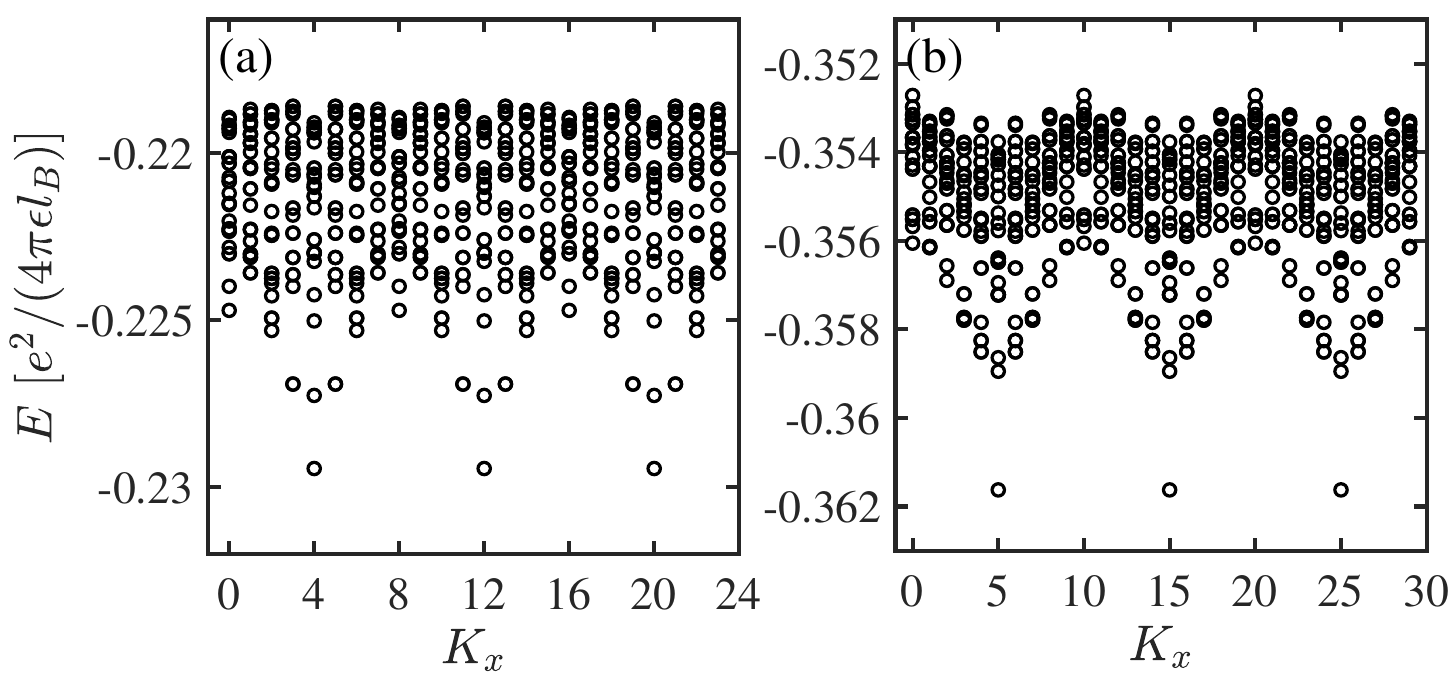}
 	\caption{
 		The many-body energy spectrum with respect to the momentum quantum number $K_x$ for
 		(a) $n_e=8$ and (b) $n_e=10$.
 		We take the aspect ratio $r_a=1$ in (a) and  $r_a=0.8$ in (b).
 	}
 	\label{Fig2}
 \end{figure}

Figure~\ref{Fig2} displays the many-body energy spectrum with respect to $K_x$,
illustrating the existence of the triply degenerate ground states for both the system sizes.
The ground states occur at the momenta $K_x=4,12,20$ for $n_e=8$ and
$K_x=5,15,25$ for $n_e=10$. The momenta are consistent with the theoretical prediction of the generalized Pauli principle for 
(1,3)-admissible partitions, that is, the maximum number of electrons in three consecutive orbitals is one~\cite{regnault2011PRX,bernevig2008PRL}.
In addition, we observe the presence of an energy gap between the ground states and the first excited states.
These results suggest that the ground states are the incompressible FQH states.
 
 \section{Ground state properties}
 
\begin{figure}[t]
	\includegraphics[width=3.4in]{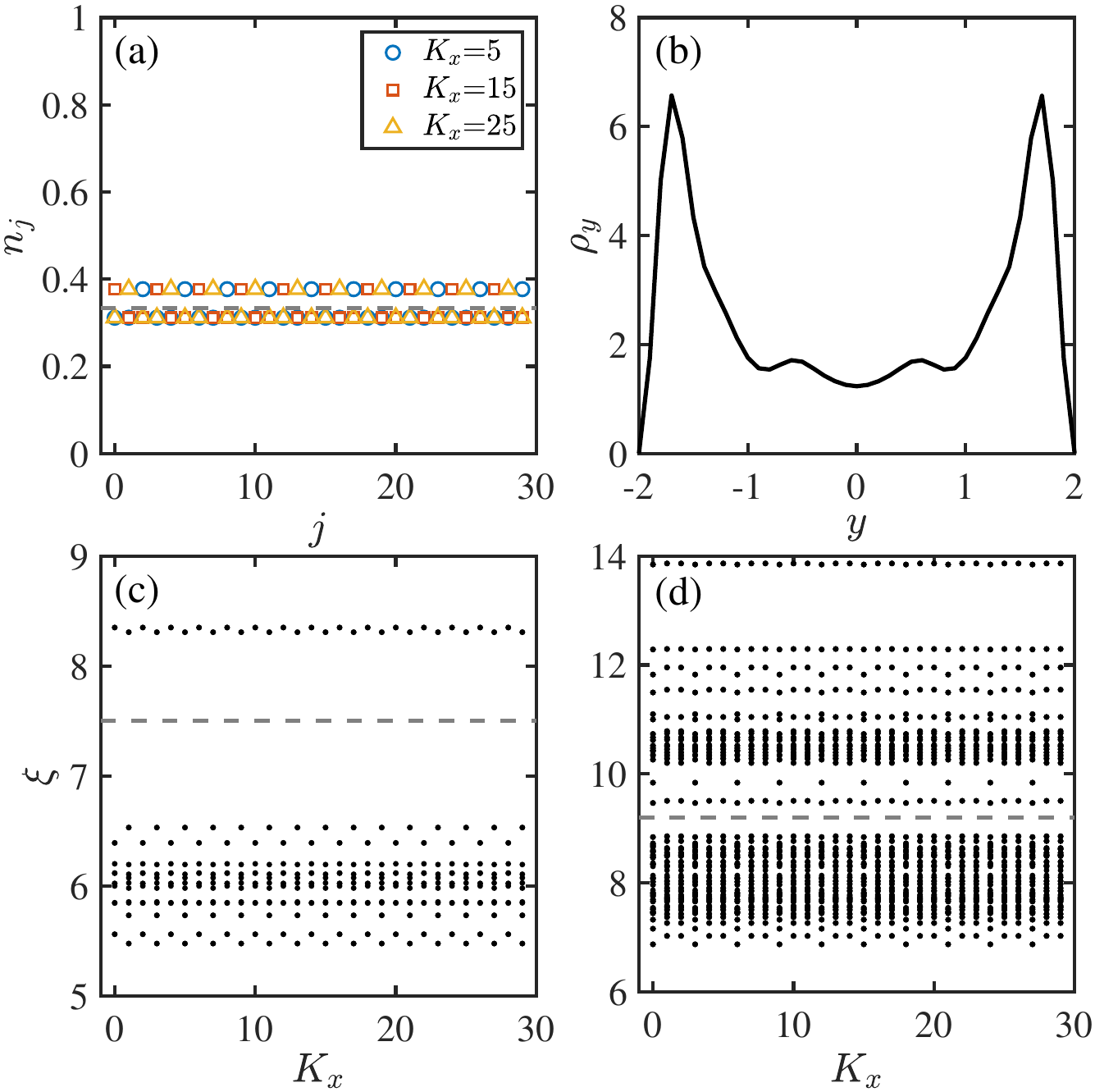}
	\caption{
		(a) The occupation number on each single-particle state with the momentum quantum number $j$ for the
		ground states with $K_x=5$ (blue circle), 10 (red square) and 15 (yellow triangle) when $n_e=10$.
		(b) The real-space electron density $\rho_y$ of the ground state with $K_x=5$
		versus the $y$ coordinate.
		The particle entanglement spectrum of the ground states for (c) $n_a=2$ and (d) $n_a=3$.
		The number of states below the gap (indicated by the dashed grey line) in each momentum sector
		is 12, 13, 12, 13, ..., 12, 13 in (c) and 85, 84, 84, 85, 84, 84, ..., 85, 84, 84 in (d),
		consistent with the (1,3) generalized Pauli principle. 
	}
	\label{Fig3}
\end{figure}

  To exclude the possibility of the CDW phase with the equal 
 ground state degeneracy, we calculate the occupation numbers on all single-particle states labelled by
the momentum quantum number $j$ for each ground state,
  $  n_j=\langle a_j^\dagger  a_j \rangle$.
As shown in Fig.~\ref{Fig3}(a), we see the occupation number on each single-particle state is always close to 1/3
(indicated by the dashed grey line) 
for every ground state of the system with $n_e=10$ electrons, 
consistent with the behavior of the FQH liquid.  
The slight deviations from the exact 1/3 should be attributed to the finite-size effect~\cite{regnault2011PRX}.   

In addition, we extract the electron number density distribution $\rho(\bm{r})$ in real space 
and obtain
the integrated density distribution $\rho_{xz}(x,z)=\int\rho(\bm{r}){\rm d} y$ in the $(x,z)$ plane,
exhibiting a quite uniform distribution with small fluctuations 
$\eta=(\max{\rho_{xz}}-\min{\rho_{xz}})/\overline{\rho}_{xz}=0.24\%$ 
where $\overline{\rho}_{xz}$ represents the average density (see Appendix D
 for the 
density distribution at a fixed $y$).
The uniformity of the electron density in both momentum and real space evidence that 
the system is in the FQH phase rather than a CDW phase.
To further investigate the spatial distribution of the electron liquid, 
we plot the integrated density distribution in the $y$ direction $\rho_{y}(y)=\int\rho(\bm{r}){\rm d}x{\rm d}z$ in Fig.~\ref{Fig3}(b).
Clearly, we see that the electron fluid concentrates around the top and bottom surfaces
where the Fermi arcs are located, 
in stark contrast to the conventional FQH effect in 2D.   
Note that the real-space electron density shown above is for the ground state with $K_x=5$,
and the results for the other two ground states are almost the same.  

 To diagnose the many-body topology of the ground states,
 we compute the Chern number by imposing the twisted boundary conditions along $x,z$ 
 ~\cite{niu1985PRB,sheng2011NC,fukui2005JPSJ} (also see Appendix E),
   		\begin{equation}\label{Chern}
 	C=\frac{1}{2\pi}\int_0^{2\pi}{\rm d}\theta_x \int_0^{2\pi} {\rm d}\theta_z{\rm Im}
 	\left [\langle {\partial_{\theta_x} \Psi} |
 	{\partial_{\theta_z} \Psi} \rangle  
 	-{\rm c.c.}\right ].	
 \end{equation}
Here $\Psi(\theta_x,\theta_z)$ is the wave function of the many-body ground state under boundary 
phases $\theta_x$ and $\theta_z$, 
which transforms $k_x$ to $k_x-\theta_x/L_x$ and $k_z$ to $k_z-\theta_z/L_z$ in the Hamiltonian 
(\ref{Ham1}).
For electron number $n_e=8$, we find that the Chern numbers of the three ground states are all 0.3321, 
very close 
to 1/3, further confirming that the ground states are the FQH states.
 
To study the excitation properties of the FQH states arising from Weyl orbits, 
we calculate the particle entanglement spectrum~\cite{regnault2011PRX,sterdyniak2011PRL} (also see Appendix F), 
that is, the eigenvalues $\xi$ of the reduced density matrix
\begin{equation}\label{ReducedDM}
    	\rho_A={\rm Tr}_B(\frac{1}{3}\sum_{i=1}^3 |\Psi_i\rangle \langle \Psi_i |) ,
\end{equation} 
where $|\Psi_i\rangle$ with $i=1,2,3$ denote the three ground states.
Here, we divide the $n_e$ electrons into $A$ and $B$ part with $n_a$ and $n_b$ electrons, 
respectively. 
Tracing out the electrons in the $B$ part corresponds to generating quasihole excitations~\cite{sterdyniak2011PRL}. 
For the total electron number $n_e=10$, we show the particle entanglement spectrum for $n_a=2$ and 
$3$
with respect to the total momentum quantum number $K_x$ in Fig.~\ref{Fig3}(c) and (d),
based on the fact that $\rho_A$ satisfies the conservation of momentum.
We see a clear gap in the entanglement spectrum for both $n_a=2$ and 3, indicated by the dashed grey
line.
For $n_a=2$, the number of states below the gap is 12 when $K_x$ is even and 13 when $K_x$ is odd;
for $n_a=3$, the number is 85 when $K_x$ is an integer multiple of 3 and 84 for other $K_x$.
The number of states in each momentum sector coincides with the analytical results of the (1,3) 
generalized Pauli principle~\cite{regnault2011PRX,bernevig2008PRL},
suggesting that the FQH states based on Weyl orbits with $\nu=1/3$ share similar excitation 
properties to the Laughlin states for the 2D electron gas. 

  \begin{figure}[t]
	\includegraphics[width=3.4in]{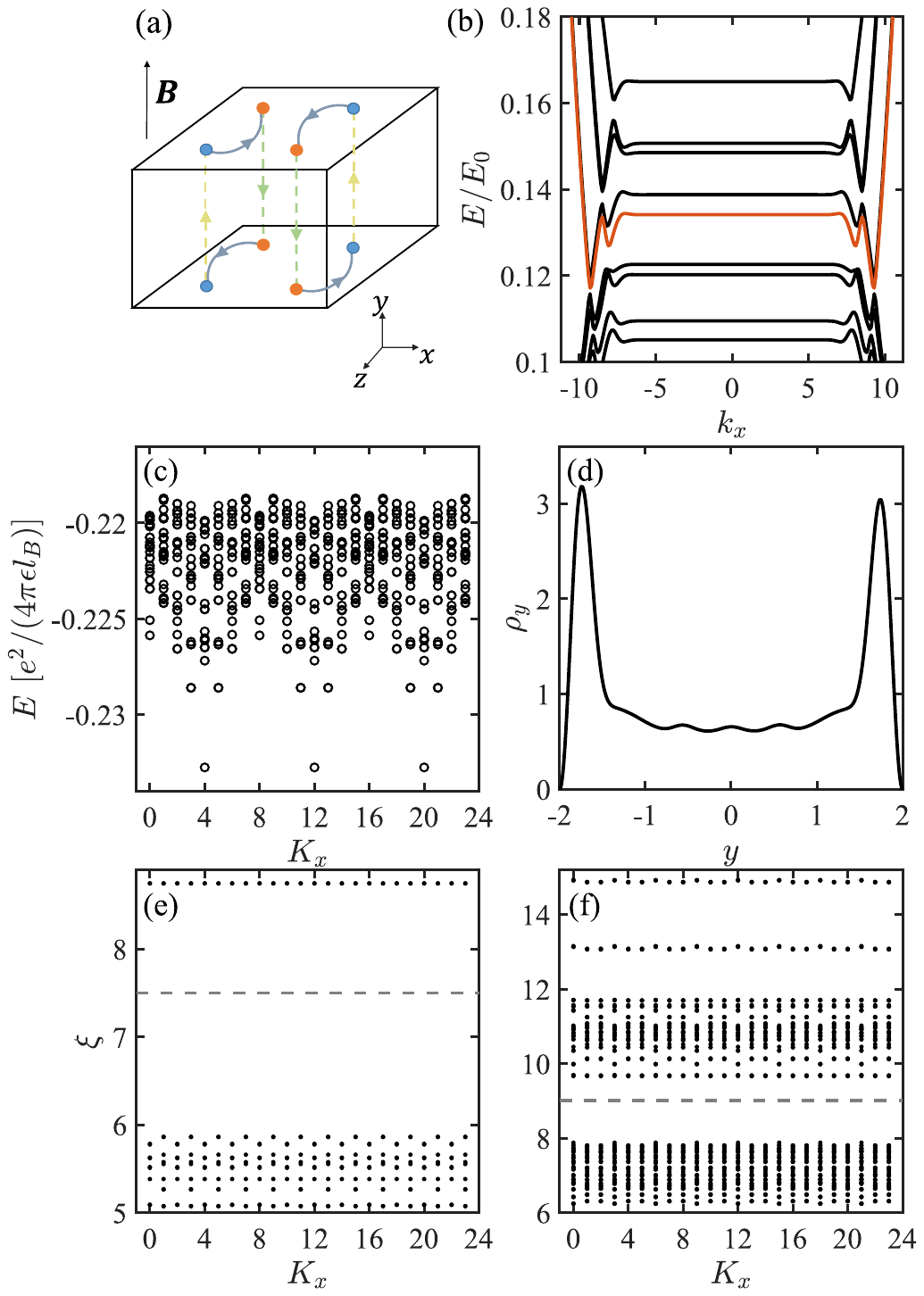}
	\caption{
		(a) Schematic illustration of the time-reversal symmetric model (\ref{Ham3}) with two pairs of 
		Weyl points, Fermi arcs and Weyl orbits formed under the magnetic field.
	(b) The Landau levels of the Hamiltonian (\ref{Ham3})
	with the partially filled Landau level colored red.
	(c) The many-body energy spectrum of the interacting Hamiltonian, exhibiting 
	the triply degenerate ground states.
	(d) The electron density distribution $\rho_y$ along $y$ of the ground state with $K_x=4$.
	Note that the density distribution is the same for the other two ground states.
	The particle entanglement spectrum of the ground states for (e) $n_a=2$ and (f) $n_a=3$.
	In (c)-(f), $n_e=8$.
	}
	\label{Fig4}
	\end{figure}

\section{Time-reversal symmetric model}

 For the generality of our results,
we study a four-band model
\begin{eqnarray}\label{Ham3}
		H_T(\bm{k})=&&M[k_w^2-(k_x^2+k_y^2+k_z^2)]s_0\sigma_z \nonumber \\ 
		&&+A(k_x s_z \sigma_x +k_ys_0\sigma_y)+\alpha k_ys_y\sigma_x \nonumber \\ 
		&&+[D_1k_y^2+D_2(k_x^2+k_z^2)]s_0\sigma_0+\beta s_y\sigma_y		
\end{eqnarray} 
with another set of identity and Pauli matrices $s_0,s_x,s_y,s_z$ representing the spin degree of 
freedom.
This model respects time-reversal symmetry, i.e. $TH_T(\bm{k})T^{-1}=H_T(-\bm{k})$ for 
the time-reversal operator $T=is_y\sigma_0\kappa$ with $\kappa$ being the complex conjugate 
operator.
The system has two pairs of Weyl points, and the projections of each pair onto the top and bottom surfaces
are connected by Fermi arcs when we take OBCs along $y$, 
as shown schematically in Fig.~\ref{Fig4}(a).
With a magnetic field $\bm{B}=(0,B,0)$ along $y$, the Fermi arcs form two Weyl orbits
assisted by the Weyl points~\cite{wang2017PRL,wang2023arXiv}.  
We plot the Landau levels near the energy of the Weyl points under PBCs along $x$
and OBCs along $y$ and $z$ in Fig.~\ref{Fig4}(b).
Here we include the Zeeman energy $H_Z=\Delta_Zs_y\sigma_0$ with 
$\Delta_Z=0.006E_0$, $\widetilde{\alpha}=\alpha/l_B=0.08E_0$ and $\beta=0.001E_0$ and other parameters the same as before.

To study the many-body physics when the Landau level we are interested in 
[colored red in Fig.~\ref{Fig4}(b)] are partially filled with $\nu=1/3$,
we project the Coulomb interaction (\ref{Coulomb}) onto that Landau level under PBCs along $x$ and $z$
and OBCs along $y$.   
In Fig.~\ref{Fig4}(c), we plot the energy spectrum obtained by diagonalizing the projected many-body Hamiltonian for
$n_e=8$, illustrating the existence of triply degenerate ground states. 
Their positions in $K_x$ are consistent with the prediction of the (1,3) 
generalized Pauli principle~\cite{bernevig2008PRL,regnault2011PRX}.
In addition,
by calculating the real-space electron density distribution $\rho(\bm{r})$ for a ground state, 
we find that the density distribution $\rho_{xz}(x,z)$ in the $(x,z)$ plane 
is almost uniform with a tiny fluctuation $\eta=0.48\%$,
characteristic of a FQH liquid behavior.
Similar to the previous case, the electron liquid inhabits the top and bottom surfaces 
[see the electron density distribution $\rho_y$ along $y$ in Fig.~\ref{Fig4}(d)], 
which constitutes the distinctive character of the FQH effect based on the Weyl orbit. 
We further calculate the particle entanglement spectrum of the ground states for $n_a=2$ and $3$,
as shown in Fig.~\ref{Fig4}(e) and (f).
We see that under the gap indicated by the dashed grey line,
the number of states is 9, 10, 9, 10, ..., 9, 10 in Fig.~\ref{Fig4}(e) and 
46, 45, 45, 46, 45, 45, ..., 46, 45, 45 in Fig.~\ref{Fig4}(f), consistent with the generalized (1,3) 
Pauli principle. All these results suggest that the ground states are the FQH states.
 
In summary, we discover the FQH effect based on Weyl orbits in Weyl semimetals under a magnetic field.
We identify the FQH effect by calculating the ground state degeneracy, the many-body Chern number, the 
density distribution and the entanglement spectrum.  
Unlike the 2D case, the FQH liquid is found to reside on two opposite surfaces of the 3D system.
Given that the 3D integer quantum Hall effect has been experimentally observed in the topological semimetal
  Cd$_3$As$_2$~\cite{zhang2017NC,uchida2017NC,schumann2018PRL,zhang2019Nature,nishihaya2021NC}, 
 we expect that the FQH effect based on Weyl orbits may
 also be observed in this or other similar materials at lower temperatures.
 Our work opens a new direction for exploring the FQH effect in 3D topological semimetals. 
 Many interesting topics, such as composite Fermi liquid and non-Abelian FQH effect,
 merit further study.

\begin{acknowledgments}
	We thank X. Gao for helpful discussions.
	This work is supported by the National Natural Science Foundation of China (Grant No. 12474265, 11974201)
	and Innovation Program for Quantum Science and Technology (Grant No. 2021ZD0301604).
	We also acknowledge the support by Center of High Performance Computing, Tsinghua University.
\end{acknowledgments}

\section{Appendix A: BASES USED TO DIAGONALIZE THE SINGLE-PARTICLE WEYL SEMIMETAL HAMILTONIAN UNDER A MAGNETIC FIELD}
In this Appendix, we will show the basis chosen to diagonalize the Weyl semimetal Hamiltonian under
the magnetic field in the main text for both open and periodic boundary conditions in the $z$ direction.	

We write the single-particle Weyl semimetal Hamiltonian (\ref{Ham1})~\cite{wang2017PRL} in the main text in the following matrix form
\begin{widetext}
\begin{equation}\label{S1Ham1}
	H=\left(\begin{array}{cc}
		(D_{2}-M)(k_{x}^{2}+k_{z}^{2})+(D_{1}-M)k_{y}^{2}+Mk_{w}^{2} & A(k_{x}-ik_{y})\\
		A(k_{x}+ik_{y}) & (D_{2}+M)(k_{x}^{2}+k_{z}^{2})+(D_{1}+M)k_{y}^{2}-Mk_{w}^{2}
	\end{array}\right).
\end{equation}
\end{widetext}
For the magnetic field in the $y$ direction $\bm{B}=(0,B,0)$, we consider the Landau gauge by 
taking the vector potential $\bm{A}=(Bz,0,0)$. As a result, $k_x$ is replaced by $k_x+z/l_B^2$ with the magnetic length 
$l_B=\sqrt{\hbar/eB}$, and $k_z$ is replaced by $-i\partial_z$ due
to the broken translational symmetry along $z$.
As the surface Fermi arc states are necessary to the Weyl orbit, we take open boundary conditions (OBCs)
along $y$ and replace $k_y$ by $-i\partial_y$.
We therefore arrive at the Hamiltonian under the magnetic field, 
\begin{widetext}
\begin{equation}\label{S1Ham2}
	H_B=\left(\begin{array}{cc}
		(D_{2}-M)[(k_{x}+z/l_B^2)^{2}-\partial ^{2}_z]-(D_{1}-M)\partial ^{2}_y+Mk_{w}^{2} & A(k_{x}+z/l_B^2-\partial_y)\\
		A(k_{x}+z/l_B^2+\partial_y) & (D_{2}+M)[(k_{x}+z/l_B^2)^{2}-\partial ^{2}_z]-(D_{1}+M)\partial ^{2}_y-Mk_{w}^{2}
	\end{array}\right)
\end{equation}
\end{widetext}
with $k_x$ being the only good quantum number.

For the calculation of Fig.~\ref{Fig1}(b) in the main text, we consider a system with a finite length $L_z$ 
and OBCs in the $z$ direction and choose the basis $|l,k\rangle$ with the real-space 
representation
\begin{equation}\label{S1Basis1}
	\begin{aligned}
	&\langle y,z|l,k\rangle= \\
	& \frac{2}{\sqrt{L_yL_z}}\sin\left[\frac{l\pi}{L_y}\left(y+\frac{L_y}{2}\right) \right]
	\sin\left[\frac{k\pi}{L_z}\left(z+\frac{L_z}{2}\right) \right]
	\end{aligned}
\end{equation}
with $l,k=1,2,3,...$ to diagonalize the Hamiltonian (\ref{S1Ham2}) such that the edge states appear in Fig.~1(b) 
where we take the cutoff $l,k\leq 60$.

If we focus on the bulk physics and do not want to see the edge states,
we consider a system that is infinite in the $z$ direction and choose the basis
$|m,h\rangle=|m\rangle\otimes|h\rangle$ with $m=0,1,2,...$ and $h=1,2,3,...$ such that 
\begin{equation}\label{S1Basis2m}
	\langle z|m\rangle=\frac{1}{\sqrt{\sqrt{\pi}2^m m!l_B}}
	\exp{\left[-\frac{(z-z_0)^2}{2l_B^2}\right]}\mathcal{H}_m(\frac{z-z_0}{l_B})
\end{equation}
with the guiding center $z_0=-k_xl_B^2$ and Hermite polynomials $\mathcal{H}_m$ and 
\begin{equation}\label{S1Basis2h}
	\langle y|h\rangle=\sqrt{\frac{2}{L_y}}\sin\left[\frac{h\pi}{L_y}\left(y+\frac{L_y}{2}\right) \right].
\end{equation}
$\langle z|m\rangle$ has the form of the wave function of the state in the $m$th Landau level in 2D 
and satisfies $\hat{b}^\dagger \hat{b}|m\rangle =m|m\rangle $,
$\hat{b}^\dagger |m\rangle =\sqrt{m+1}|m+1\rangle $ and $\hat{b}|m\rangle =\sqrt{m}|m-1\rangle $
where
\begin{equation}\label{S1Ope1}
	\hat{b}^\dagger=\frac{l_B}{\sqrt{2}}\left[k_x+\frac{z}{l_B^2}-i(-i\partial_z)\right],
\end{equation}
\begin{equation}\label{S1Ope2}
	\hat{b}=\frac{l_B}{\sqrt{2}}\left[k_x+\frac{z}{l_B^2}+i(-i\partial_z)\right].
\end{equation}
In terms of $\hat{b}^\dagger$ and $\hat{b}$, the Hamiltonian (\ref{S1Ham2}) is expressed as
\begin{widetext}
\begin{equation}\label{S1Ham3}
	H_B=\left(\begin{array}{cc}
		\frac{2(D_2-M)}{l_B^2}(\hat{b}^\dagger \hat{b}+\frac{1}{2})-(D_{1}-M)\partial ^{2}_y+Mk_{w}^{2} & A[\frac{\hat{b}^\dagger +\hat{b}}{\sqrt{2}l_B}-\partial_y]\\
		A[\frac{\hat{b}^\dagger +\hat{b}}{\sqrt{2}l_B}+\partial_y]& 	\frac{2(D_2+M)}{l_B^2}(\hat{b}^\dagger \hat{b}+\frac{1}{2})-(D_{1}+M)\partial ^{2}_y-Mk_{w}^{2}
	\end{array}\right).
\end{equation}
\end{widetext}
By diagonalizing the Hamiltonian (\ref{S1Ham3}) under the basis $|m,h\rangle$ ,
we get the $u$th eigenstate 
\begin{equation}\label{S1State1}
	|u\rangle=\sum_{m,h,\sigma}c_{m,h,\sigma}^u|m,h\rangle\otimes|\sigma \rangle.
\end{equation}
with $\sigma=0,1$ representing the pseudospin degree of freedom 
and the real-space wave function is the spinor
\begin{equation}\label{S1WF0}
	\bm{f^u}=(f_0^u,f_1^u)^T 	 
\end{equation}
with   	 $f_\sigma^u(y,z)=\langle y,z,\sigma	|u\rangle$.
Note that the diagonalization process is performed in the subspace with a specific momentum $k_x$,
so the complete real-space wave function of the corresponding eigenstate is $ \bm{\tilde{\phi}^u}=(\tilde{\phi}^u_0,\tilde{\phi}^u_1)^T$ where
\begin{equation}\label{S1WF1}
	\begin{aligned}
		\tilde{\phi}^u_\sigma(x,y,z)&=e^{ik_xx}f_\sigma^u(y,z)\\
		&=\sum_{m,h}c_{m,h,\sigma}^u\tilde{\psi}_{m,h}(x,y,z)
	\end{aligned}
\end{equation}
with
\begin{eqnarray}\label{S1Basis3}
	\tilde{\psi}_{m,h}(x,y,z)&=&\sqrt{\frac{2}{\sqrt{\pi}2^m m!l_BL_y}}
	\exp{\left[ik_xx-\frac{(z-z_0)^2}{2l_B^2}\right]}\nonumber\\
&&	\mathcal{H}_m(\frac{z-z_0}{l_B})
	\sin\left[\frac{h\pi}{L_y}\left(y+\frac{L_y}{2}\right) \right].
\end{eqnarray}

In the main text, when we project the Coulomb interaction onto a Landau level,
we consider the system with the finite length $L_x,L_z$ and periodic boundary conditions (PBCs) in the $x,z$ directions.
In this case, the eigenstate wave function is  $(\phi^u_{0,j},\phi^u_{1,j})^T$ where
\begin{equation}\label{S1WF2}
	\phi^u_{\sigma,j}(x,y,z)=\sum_{m,h}c_{m,h,\sigma}^u\psi_{m,h,j}(x,y,z)
\end{equation}
with 
\begin{eqnarray}\label{S1Basis4}
	\psi_{m,h,j}(x,y,z)&=&\sqrt{\frac{2}{\sqrt{\pi}2^m m!l_BL_y}}\nonumber\\
	&&\sum_{n=-\infty}^{+\infty}\exp{\left[i(\frac{2\pi j}{L_x}+\frac{nL_z}{l_B^2})x-\frac{(z-z_n)^2}{2l_B^2}\right]}\nonumber\\
	&&\mathcal{H}_m(\frac{z-z_n}{l_B})
	\sin\left[\frac{h\pi}{L_y}\left(y+\frac{L_y}{2}\right) \right].
\end{eqnarray}
Here, $z_n=-2\pi jl_B^2/L_x-nL_z$ where the momentum quantum number $j=0,1,2,...,m_L-1$ with $m_L=L_xL_z/(2\pi l_B^2)$ being the Landau level degeneracy.
The basis wave function satisfies $\psi_{m,h,j}(x,y,z+L_z)=e^{-iL_zx/l_B^2}\psi_{m,h,j}(x,y,z)$
and $\psi_{m,h,j}(x+L_x,y,z)=\psi_{m,h,j}(x,y,z)$, which is consistent with the magnetic translation
operator $\mathcal{T}_{\bm R}$ acting as $\mathcal{T}_{\bm R}\psi ({\bm r})=e^{ixR_z/l_B^2}\psi ({\bm r}+{\bm R})$ under the gauge vector $\bm{A}=(Bz,0,0)$
with ${\bm R}=(R_x,0,R_z)$ being the translation vector.

\section{Appendix B: PROPERTIES OF THE LANDAU LEVEL BASED ON WEYL ORBITS}
In this Appendix, we will show the weight of the considered Landau level based on 
Weyl orbits in the wave functions in the considered basis so as to validate the cutoff used for 
the calculation in the main text and provide the real-space 
wave function to illustrate that the Landau level based on Weyl orbits is very different from 
the lowest Landau level in 2D.
\begin{figure}
	\includegraphics[width=3.4in]{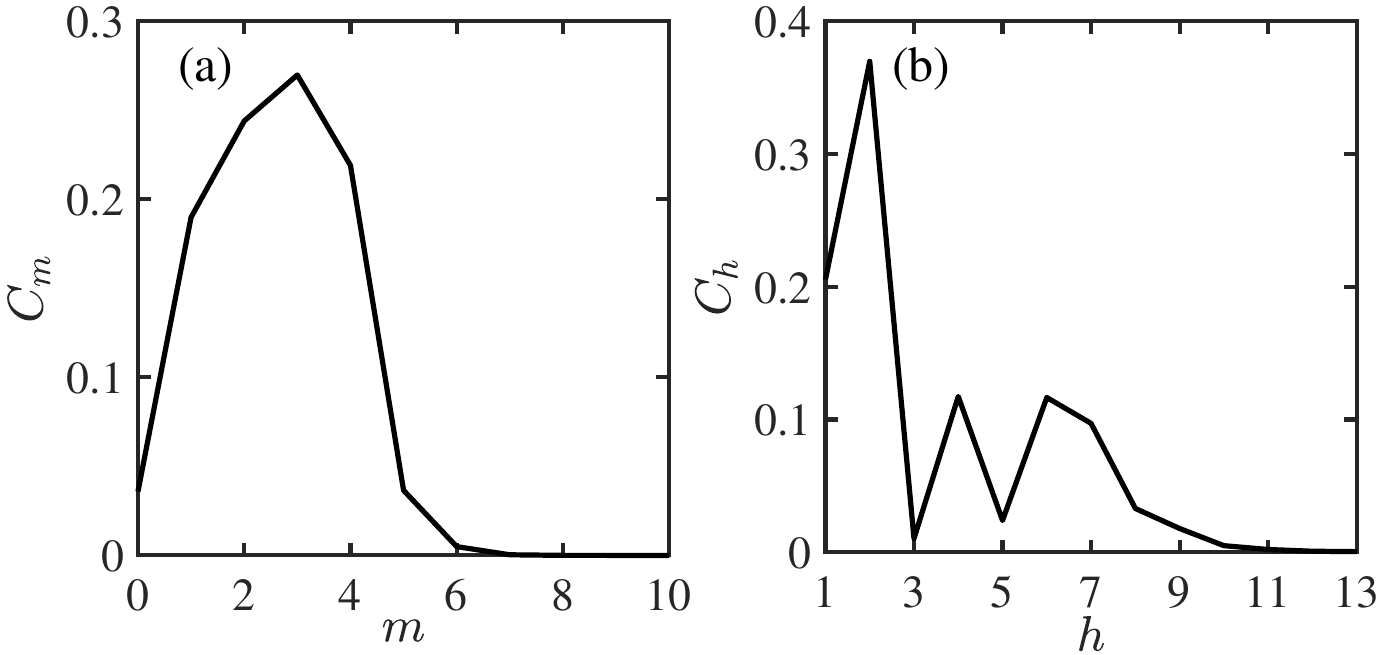}
	\caption{
		The probability (a) $C_m$ and (b) $C_h$ for a state in the considered Landau level in
		the main text [colored red in Fig.~\ref{Fig1}(b)].	}
	\label{FigS1}
\end{figure}

We first show the probability that an eigenstate in the considered Landau level in the main text is
found at different $m$ and $h$,
\begin{equation}\label{S2Probm}
	C_m=\sum_{h,\sigma}|c^u_{m,h,\sigma}|^2,
\end{equation}
and
\begin{equation}\label{S2Probh}
	C_h=\sum_{m,\sigma}|c^u_{m,h,\sigma}|^2.
\end{equation} 	 
In Fig.~\ref{FigS1}(a), we see that the maximum value of $C_m$ locates at $m=3$ and the value at $m=0$
is relatively small compared with that at $m=1,2,3,4$, 
For larger $m$, $C_m$ quickly drops to 0, so we take the cutoff $m\leq 8$ for the calculation of the 
interacting Hamiltonian (3) in the main text.  
With the increasing of $h$, $C_h$ reaches the maximum at $h=2$ and then undergoes a sharp 
drop, followed by an oscillation before decaying to 0, as shown in Fig.~\ref{FigS1}(b).
We thus take the cutoff $h\leq 13$ in the main text.
To justify the validity of the cutoff, we also calculate the many-body spectrum with larger $m$ and $h$
and find that the spectrum almost does not change up to $m=13$ and $h=20$, 
revealing that the above cutoff used in the main text is enough.

\begin{figure*}
	\includegraphics[width=6in]{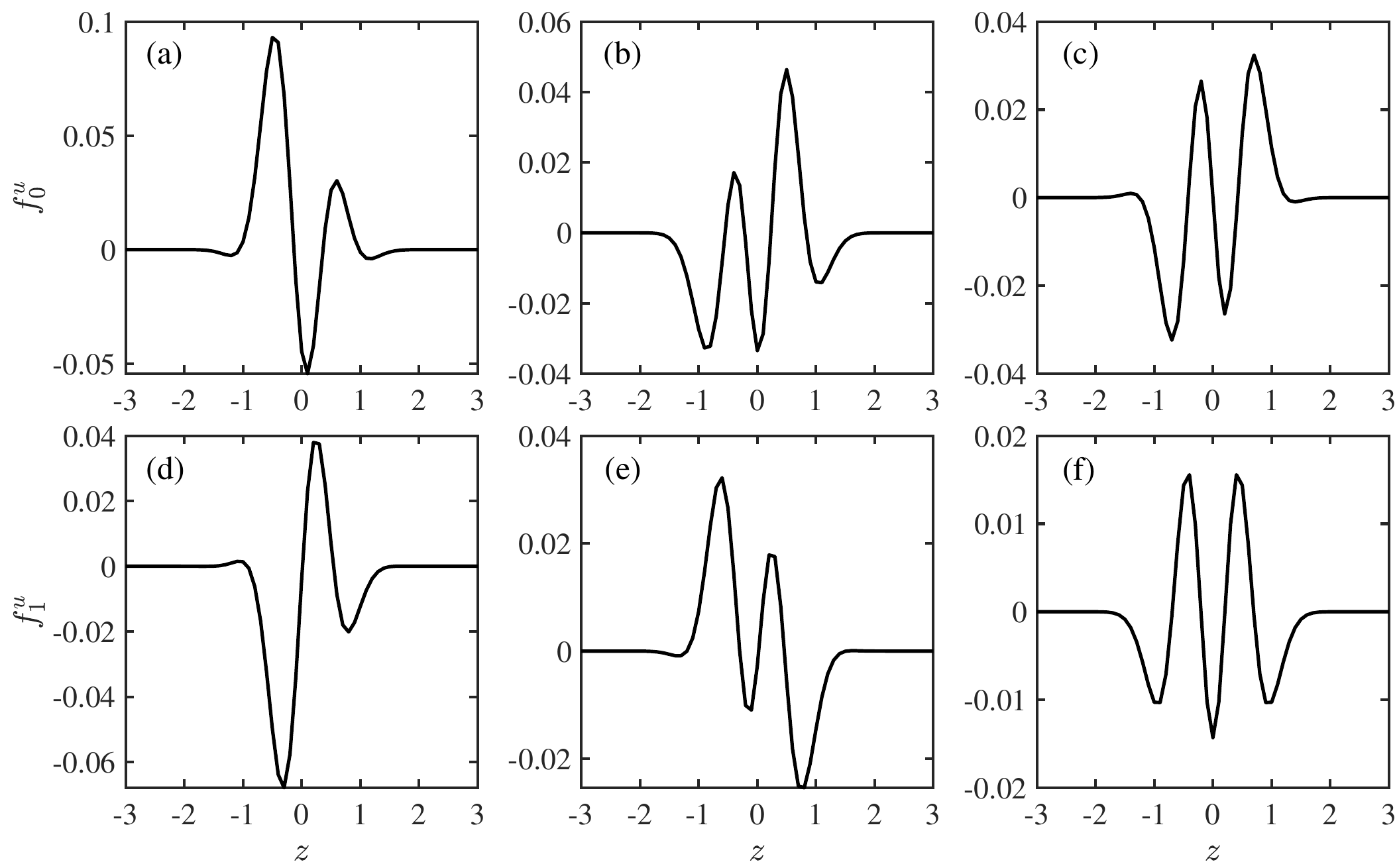}
	\caption{
		The wave function versus $z$ [the $f_0^u$ ($f_1^u$) component in the first (second) row] of the considered Landau level state based on Weyl orbits
		at a fixed $y$. In (a) and (d), $y=-1.7$, in (b) and (e), $y=-1.1$, and in (c) and (f), $y=0$.   }
	\label{FigS2}
\end{figure*}

To show that the Landau level based on Weyl orbits is very different from the lowest Landau level
in 2D, we plot the wave function $\bm{f^u}$ in Eq.~(\ref{S1WF0}) versus $z$ at certain $y$ for a system with thickness $L_y=4$ and $y\in[-L_y/2,L_y/2]$ in Fig.~\ref{FigS2}.
Around the surface, as shown in Fig.~\ref{FigS2}(a) for $y=-1.7$, the component $f_0^u$ exhibits two nodes if we neglect the tiny dip near $z=\pm 1$.
Going inside into the bulk, $f_0^u$ hosts four nodes at $y=-1.1$ [see Fig.~\ref{FigS2}(b)] and three nodes at $y=0$ [see Fig.~\ref{FigS2}(c)].
For the other component $f_1^u$, the numbers of nodes are 2, 3, 4 at $y=-1.7,-1.1,0$, as shown in
Fig.~\ref{FigS2}(d),(e) and (f).
We thus see that the Laudau level based on Weyl orbits is distinct from the lowest Landau level in 2D 
which carries no nodes. It is therefore not immediately apparent that the FQH effect can also occur in this system.

\section{Appendix C: THE COEFFICIENTS OF THE INTERACTING HAMILTONIAN}
In this Appendix, we will derive the coefficients $B_{j_1j_2j_3j_4}$ of the interacting Hamiltonian (3) in the
main text in detail.
The interacting Hamiltonian in the second quantization form reads
\begin{align}\label{S3Ham1}
	\hat{H}_I=\frac{1}{2}\int {\rm d}\bm{r}_{1}\int {\rm d}\bm{r}_{2}\sum_{\sigma_1,\sigma_2}\Psi^{\dagger}_{\sigma_1}(\bm{r}_{1})\Psi^{\dagger}_{\sigma_2}(\bm{r}_{2})\nonumber\\
	V(\bm{r}_{1}-\bm{r}_{2})\Psi_{\sigma_2}(\bm{r}_{2})\Psi_{\sigma_1}(\bm{r}_{1}),
\end{align}
where $\Psi^{\dagger}_{\sigma}(\bm{r})$ and $\Psi_{\sigma}(\bm{r})$ are creation and annihilation field operators with
the pseudospin degree of freedom $\sigma$,
\begin{equation}\label{S3Coulomb}
	V(\bm{r})=\sum_{t,s=-\infty}^{+\infty}\frac{e^2}{4\pi\epsilon}\frac{1}{\bm{r}+tL_x\bm{e_x}+
		sL_z\bm{e_z}}	
\end{equation} 
is the Coulomb interaction under PBCs in the $x$ and $z$ directions, and 
$\int {\rm d}{\bm r}:= \int_0^{L_x} {\rm d}x \int_{-L_y/2}^{L_y/2}{\rm d}y\int_{-L_z}^0 {\rm d}z$
for a system with the length $L_x,L_y,L_z$ in the $x,y,z$ directions, respectively.
Under the approximation of the single band projection,
\begin{equation}\label{S3FiledOp1}
	\Psi_{\sigma}^{\dagger}(\bm{r})=\sum_{j=0}^{m_{L}-1}\phi^{*u}_{\sigma,j}(\bm{r})a_{j}^{\dagger}
\end{equation} 
and
\begin{equation}\label{S3FiledOp2}
	\Psi_{\sigma}(\bm{r})=\sum_{j=0}^{m_{L}-1}\phi^{u}_{\sigma,j}(\bm{r})a_{j}
\end{equation} 
where $\phi^{u}_{\sigma,j}$ is the wave function of the state with the momentum quantum number $j$ in the 
Landau level that we project the Coulomb interaction onto and 
$\hat{a}^\dagger_{j_\mu}$ ($\hat{a}_{j_\mu}$) is the corresponding creation (annihilation) operator. 
Substituting Eq.~(\ref{S3FiledOp1}) and Eq.~(\ref{S3FiledOp2}) into Eq.~(\ref{S3Ham1}) and using Eq.~(\ref{S1WF2}),
we get
\begin{align}\label{S3Ham2}
	\hat{H}_I&= \sum_{j_1j_2j_3j_4}\frac{1}{2}\int {\rm d}\bm{r}_{1}\int {\rm d}\bm{r}_{2}
	\sum_{\sigma_1,\sigma_2}\phi^{*u}_{\sigma_1,j_1}(\bm{r_1})
	\phi^{*u}_{\sigma_2,j_2}(\bm{r_2})\nonumber\\
	&V(\bm{r}_{1}-\bm{r}_{2})
	\phi^{u}_{\sigma_2,j_3}(\bm{r_2})\phi^{u}_{\sigma_1,j_4}(\bm{r_1})
	\hat{a}^\dagger_{j_1}\hat{a}^\dagger_{j_2}
	\hat{a}_{j_3}\hat{a}_{j_4},
\end{align}
so the coefficient in Eq.~(\ref{Ham2}) in the main text is
	\begin{align}\label{S3B}
		B_{j_1j_2j_3j_4}&=\frac{1}{2}\int {\rm d}\bm{r}_{1}\int {\rm d}\bm{r}_{2}
		\sum_{\sigma_1,\sigma_2}\phi^{*u}_{\sigma_1,j_1}(\bm{r_1})
		\phi^{*u}_{\sigma_2,j_2}(\bm{r_2})\nonumber\\
		&V(\bm{r}_{1}-\bm{r}_{2})
		\phi^{u}_{\sigma_2,j_3}(\bm{r_2})\phi^{u}_{\sigma_1,j_4}(\bm{r_1})\nonumber\\
		&=\sum_{\sigma_1\sigma_2}\sum_{m_{1,2,3,4}h_{1,2,3,4}}c_{m_1,h_1,\sigma_1}^{*u}
		c_{m_2,h_2,\sigma_2}^{*u}\nonumber\\
		&c_{m_3,h_3,\sigma_2}^{u}c_{m_4,h_4,\sigma_1}^{u}
		A_{j_1j_2j_3j_4}^{m_1h_1m_2h_2m_3h_3m_4h_4}
	\end{align}
where 
\begin{align}\label{S3A}
	A_{j_1j_2j_3j_4}^{m_1h_1m_2h_2m_3h_3m_4h_4}=\frac{1}{2}\int {\rm d}\bm{r}_{1}\int {\rm d}\bm{r}_{2}
	\psi_{m_1h_1j_1}^*({\bm r_1})\nonumber\\
	\psi_{m_2h_2j_2}^*({\bm r_2})
	V(\bm{r}_{1}-\bm{r}_{2})
	\psi_{m_3h_3j_3}({\bm r_2})\psi_{m_4h_4j_4}({\bm r_1}).
\end{align}
To evaluate $A_{j_1j_2j_3j_4}^{m_1h_1m_2h_2m_3h_3m_4h_4}$,
we make use of the explicit form of $\psi_{m,h,j}({\bm r})$ in Eq.~(\ref{S1Basis4})
and the Fourier transform in the $x$ and $z$ directions of the Coulomb interaction
\begin{equation}\label{S3CouFourier}
	V(q_x,q_z,y)=\frac{1}{L_xL_z}\frac{2\pi e^{-q|y|}}{q},
\end{equation}
where $q=\sqrt{q_x^2+q_z^2}$ such that
\begin{equation}\label{S3Coulomb2}
	V({\bm r})=\sum_{q_x,q_z,s,t}^\prime V({\bm q},y)e^{iq_xx+iq_zz}\delta_{q_x,\frac{2\pi}{L_x}t}
	\delta_{q_z,\frac{2\pi}{L_z}s}
\end{equation}
with ${\bm q}=(q_x,q_z)$ and $s,t$ taking integer values from $-\infty$ to $+\infty$ excluding 0,
indicated by the symbol $^\prime$ on the summation symbol.
We can safely drop the coefficient with ${\bm q}={\bm 0}$ because it becomes a constant proportional to
the electron density, which will be cancelled by the positive charge background in the thermodynamic limit.
In a finite system, it only contributes a constant to the diagonal elements of the many-body Hamiltonian
and does not change the physics, see, e.g. the Supplemental Material in Ref.~\cite{reddy2023PRB}.    
Then
\begin{align}\label{S3A2}
	A_{j_1j_2j_3j_4}^{m_1h_1m_2h_2m_3h_3m_4h_4}=\frac{1}{2L_xL_z}
	\sum_{q_x,q_z,s,t}^\prime \frac{2\pi}{q}F({\bm q})\nonumber\\
	S_1({\bm q})S_2(\bm q)\delta_{q_x,\frac{2\pi}{L_x}t}
	\delta_{q_z,\frac{2\pi}{L_z}s},
\end{align}
where 
\begin{align}\label{S3F}
	F(\bm{q})=&\int_{-L_y/2}^{L_y/2} \int_{-L_y/2}^{L_y/2}{\rm d}y_{1}{\rm d}y_{2}(\frac{2}{L_{y}})^{2}\sin[\frac{h_{1}\pi}{L_{y}}(y_{1}+\frac{L_{y}}{2})]\nonumber\\
	&\sin[\frac{h_{4}\pi}{L_{y}}(y_{1}+\frac{L_{y}}{2})]\sin[\frac{h_{2}\pi}{L_{y}}(y_{2}+\frac{L_{y}}{2})]\nonumber\\
	&\sin[\frac{h_{3}\pi}{L_{y}}(y_{2}+\frac{L_{y}}{2})]e^{-q|y_{1}-y_{2}|},
\end{align}
	\begin{align}\label{S3S1}
		S_1({\bm q})=a\sum_{n_1n_4}\int_0^{L_x}{\rm d}x_1\frac{1}{L_x}e^{-i\left[\frac{2\pi}{L_x}(j_1-j_4)+\frac{(n_1-n_4)L_z}{l_B^2}-q_x\right]
			x_1	}&\nonumber\\
		\int_{-L_z}^0 {\rm d}z_{1}e^{iq_{z}z_{1}}e^{-[(Z_{j_{1}}+n_{1}L_z+z_{1})^{2}+(Z_{j_{4}}+n_{4}L_z+z_{1})^{2}]/(2l_B^{2})}&\nonumber 	\\
		H_{m_{1}}[(z_{1}+Z_{j_{1}}+n_{1}L_z)/l_B]H_{m_{4}}[(z_{1}+Z_{j_{4}}+n_{4}L_z)/l_B],&
	\end{align}	
	\begin{align}\label{S3S2}
		S_2({\bm q})=a\sum_{n_2n_3}\int_0^{L_x}{\rm d}x_2\frac{1}{L_x}e^{-i\left[\frac{2\pi}{L_x}(j_2-j_3)+\frac{(n_2-n_3)L_z}{l_B^2}+q_x\right]
			x_2	}\nonumber\\
		\int_{-L_z}^0 {\rm d}z_{2}e^{iq_{z}z_{2}}e^{-[(Z_{j_{2}}+n_{2}L_z+z_{2})^{2}+(Z_{j_{3}}+n_{3}L_z+z_{2})^{2}]/(2l_B^{2})} \nonumber	\\
		H_{m_{2}}[(z_{2}+Z_{j_{2}}+n_{2}L_z)/l_B]H_{m_{3}}[(z_{2}+Z_{j_{3}}+n_{3}L_z)/l_B]
	\end{align}
with $Z_j=2\pi jl_B^2/L_x,a=1/(l_B\sqrt{\pi}\sqrt{2^{m_{1}+m_{4}}m_{1}!m_{4}!})$.
In the following, we derive $S_1({\bm q})$, and some explanations for certain steps can be found after Eq.~(\ref{S1_7}).
\begin{widetext}
\begin{align}
	S_1({\bm q})&=a\sum_{\Delta n}\delta_{q_x,\frac{2\pi}{L_x}[j_1-j_4+\Delta nm_L]}
	\int_{-L_z}^0 	\sum_n{\rm d}z_{1}e^{iq_{z}z_{1}}e^{-[(Z_{j_{1}}+\Delta nL_z+z_{1}+nL_z)^{2}+(Z_{j_{4}}+z_{1}+nL_z)^{2}]/(2l_B^{2})}\notag\\
	&H_{m_{1}}[(z_{1}+nL_z+Z_{j_{1}}+\Delta nL_z)/l_B]H_{m_{4}}[(z_{1}+nL_z+Z_{j_{4}})/l_B]\label{S1_1}\\
	&=a\sum_{\Delta n}\delta_{q_x,\frac{2\pi}{L_x}[j_1-j_4+\Delta nm_L]}
	\int_{-\infty}^{+\infty}{\rm d}z_{1}e^{iq_{z}z_{1}}e^{-[(Z_{j_{1}}+\Delta nL_z+z_{1})^{2}+(Z_{j_{4}}+z_{1})^{2}]/(2l_B^{2})}\notag\\
	&H_{m_{1}}[(z_{1}+Z_{j_{1}}+\Delta nL_z)/l_B]H_{m_{4}}[(z_{1}+Z_{j_{4}})/l_B]\label{S1_2}\\
	&=a\sum_{\Delta n}\delta_{q_x,\frac{2\pi}{L_x}[j_1-j_4+\Delta nm_L]}
	\lim_{t_1,t_4\rightarrow 0}\frac{{\rm d}^{m_1}}{{\rm d}t_1^{m_1}}
	\frac{{\rm d}^{m_4}}{{\rm d}t_4^{m_4}}
	\int_{-\infty}^{+\infty}{\rm d}z_{1}e^{iq_{z}z_{1}}e^{-[(Z_{j_{1}}+\Delta nL_z+z_{1})^{2}+(Z_{j_{4}}+z_{1})^{2}]/(2l_B^{2})}\notag\\
	&e^{-t_1^2+2t_1(z_{1}+Z_{j_{1}}+\Delta nL_z)/l_B-t_4^2+2t_4(z_{1}+Z_{j_{4}})/l_B}\label{S1_3}\\
	&=a\sum_{\Delta n}\delta_{q_x,\frac{2\pi}{L_x}[j_1-j_4+\Delta nm_L]}
	\lim_{t_1,t_4\rightarrow 0}\frac{{\rm d}^{m_1}}{{\rm d}t_1^{m_1}}
	\frac{{\rm d}^{m_4}}{{\rm d}t_4^{m_4}}
	e^{\frac{-[(Z_{j_{1}}+\Delta n L_z)^{2}+Z_{j_{4}}^{2}]}{2l_B^{2}}+2\frac{Z_{j_{1}}+
			\Delta nL_z}{l_B}t_{1}+2\frac{Z_{j_{4}}}{l_B}t_{4}-t_{1}^{2}-t_{4}^{2}}\notag\\
	& \int_{-\infty}^{+\infty}{\rm d}z_{1}e^{iq_{z}z_{1}}
	e^{[-z_{1}^{2}-(Z_{j_{1}}+\Delta nL_z+Z_{j_{4}}-2t_{1}l_B-2t_{4}l_B)x_{1}]/l_B^{2}}\label{S1_4}\\
	&=a^\prime \sum_{\Delta n}\delta_{q_x,\frac{2\pi}{L_x}[j_1-j_4+\Delta nm_L]}
	\lim_{t_1,t_4\rightarrow 0}\frac{{\rm d}^{m_1}}{{\rm d}t_1^{m_1}}
	\frac{{\rm d}^{m_4}}{{\rm d}t_4^{m_4}}
	e^{\frac{-[(Z_{j_{1}}+\Delta n L_z)^{2}+Z_{j_{4}}^{2}]}{2l_B^{2}}+2\frac{Z_{j_{1}}+
			\Delta nL_z}{l_B}t_{1}+2\frac{Z_{j_{4}}}{l_B}t_{4}-t_{1}^{2}-t_{4}^{2}
		+\gamma^2/l_B^2-q_z^2l_B^2/4+iq_z\gamma}\label{S1_5}\\
	&=a^\prime \sum_{\Delta n}\delta_{q_x,\frac{2\pi}{L_x}[j_1-j_4+\Delta nm_L]}
	e^{-q_z^2l_B^2/4-[(Z_{j_{1}}+\Delta nL_z)-Z_{j_{4}}]^{2}/(4l_B^2)
		-iq_{z}(Z_{j_{1}}+\Delta nL_z+Z_{j_{4}})/2}\notag\\
	&\lim_{t_1,t_4\rightarrow 0}\frac{{\rm d}^{m_1}}{{\rm d}t_1^{m_1}}
	\frac{{\rm d}^{m_4}}{{\rm d}t_4^{m_4}}
	e^{[(Z_{j_{1}}+\Delta nL_z)-Z_{j_{4}}]t_1/l_B-[(Z_{j_{1}}+\Delta nL_z)-Z_{j_{4}}]t_4/l_B
		+2t_1t_4+i(t_1+t_4)q_zl_B}\label{S1_6}\\
	&=a^\prime \sum_{\Delta n}\delta_{q_x,\frac{2\pi}{L_x}[j_1-j_4+\Delta nm_L]}
	e^{-q^2l_B^2/4-iq_{z}(Z_{j_{1}}+\Delta nL_z+Z_{j_{4}})/2}
	\lim_{t_1,t_4\rightarrow 0}\frac{{\rm d}^{m_1}}{{\rm d}t_1^{m_1}}
	\frac{{\rm d}^{m_4}}{{\rm d}t_4^{m_4}}
	e^{q_xl_B(t_1-t_4)+2t_1t_4+i(t_1+t_4)q_zl_B}.\label{S1_7}
\end{align}
\end{widetext}
From Eq.~(\ref{S3S1}) to Eq.~(\ref{S1_1}), we use 
$n=n_4,\Delta n=n_1-n_4$ 
and $m_L=L_xL_z/(2\pi l_B^2)$.
From Eq.~(\ref{S1_1}) to Eq.~(\ref{S1_2}), we utilize the Poisson summation rule.
We use the generating function of the Hermite polynomial
\begin{equation}\label{S3Hermite}	
	\frac{{\rm d}^{m}}{{\rm d}t^{m}}e^{-t^2+2tx}=H_m(x)
\end{equation}
to derive from Eq.~(\ref{S1_2}) to Eq.~(\ref{S1_3}).
From Eq.~(\ref{S1_4}) to Eq.~(\ref{S1_5}), we use the complex form Gaussian integral
\begin{equation}\label{Gaussian}
	\int_{-\infty}^{+\infty} dxe^{iqx}e^{-x^{2}+sx}=\sqrt{\pi}e^{(s^{2}-q^{2})/4+iqs/2}
\end{equation}
and $a^\prime=al_B\sqrt(\pi)=1/(2^{m_{1}+m_{4}}m_{1}!m_{4}!)$ as well as
$\gamma=-(Z_{j_{1}}+\Delta nL_z+Z_{j_{4}}-2t_{1}l_B-2t_{4}l_B)/2$.
We use $[Z_{j_{1}}+\Delta nL_z)-Z_{j_{4}}]/l_B^2=q_x$ due to
$\delta_{q_x,2\pi[j_1-j_4+\Delta nm_L]/L_x}$
to arrive at Eq.~(\ref{S1_7}). 

When $m_1\leq m_4$,
\begin{widetext}
\begin{align}
	S_1({\bm q})&=\frac{1}{\sqrt{2^{m_{1}+m_{4}}m_{1}!m_{4}!}}\sum_{\Delta n}\delta_{q_x,\frac{2\pi}{L_x}[j_1-j_4+\Delta nm_L]}
	e^{-q^2l_B^2/4-iq_{z}(Z_{j_{1}}+\Delta nL_z+Z_{j_{4}})/2}\notag\\
	& \lim_{t_1,t_4\rightarrow 0}\frac{{\rm d}^{m_1}}{{\rm d}t_1^{m_1}}
	\frac{{\rm d}^{m_4}}{{\rm d}t_4^{m_4}}
	e^{(2t_1-q_xl_B+iq_zl_B)t_4+(q_xl_B+iq_zl_B)t_1}\label{S1_8}\\
	&=\frac{2^{m_4}}{\sqrt{2^{m_{1}+m_{4}}m_{1}!m_{4}!}}\sum_{\Delta n}\delta_{q_x,\frac{2\pi}{L_x}[j_1-j_4+\Delta nm_L]}
	e^{-q^2l_B^2/4-iq_{z}(Z_{j_{1}}+\Delta nL_z+Z_{j_{4}})/2}\notag\\
	&\lim_{t_1\rightarrow 0}\frac{{\rm d}^{m_1}}{{\rm d}t_1^{m_1}}
	(t_1-q_xl_B/2+iq_zl_B/2)^{m_4}e^{(q_xl_B+iq_zl_B)t_1}\label{S1_9}\\
	&=\frac{2^{m_4}}{\sqrt{2^{m_{1}+m_{4}}m_{1}!m_{4}!}}\sum_{\Delta n}\delta_{q_x,\frac{2\pi}{L_x}[j_1-j_4+\Delta nm_L]}
	e^{-q^2l_B^2/4-iq_{z}(Z_{j_{1}}+\Delta nL_z+Z_{j_{4}})/2}\notag\\
	&\sum_{u=0}^{m_{1}}[(q_{x}+iq_{z})l_B]^{m_{1}-u}
	[\frac{(-q_{x}+iq_{z})l_B}{2}]^{m_{4}-u}
	\frac{m_{4}!}{(m_{4}-u)!}\frac{m_{1}!}{(m_{1}-u)!u!}\label{S1_10}\\
	&=\sqrt{\frac{m_1!}{m_4!}2^K}
	\sum_{\Delta n}\delta_{q_x,\frac{2\pi}{L_x}[j_1-j_4+\Delta nm_L]}
	e^{-q^2l_B^2/4-iq_{z}(Z_{j_{1}}+\Delta nL_z+Z_{j_{4}})/2}\notag\\
	&[\frac{(-q_{x}+iq_{z})l_B}{2}]^{K}
	\sum_{s=0}^{m_{1}}(-1)^{s}
	(\frac{q^{2}l_B^{2}}{2})^{s}\frac{m_{4}!}{(s+K)!s!(m_{1}-s)!}\label{S1_11}\\
	&=\sqrt{\frac{m_1!}{m_4!}}
	\sum_{\Delta n}\delta_{q_x,\frac{2\pi}{L_x}[j_1-j_4+\Delta nm_L]}
	e^{-q^2l_B^2/4-iq_{z}(Z_{j_{1}}+\Delta nL_z+Z_{j_{4}})/2}
	[\frac{(-q_{x}+iq_{z})l_B}{\sqrt{2}}]^{K}\mathcal{L}_{m_1}^{(K)}(\frac{q^2l_B^2}{2}).
	\label{S1_12}
\end{align}
\end{widetext}
From Eq.~(\ref{S1_9}) to Eq.~(\ref{S1_10}), we use the formula
\begin{align}\label{iterative}
	\frac{d^{n}}{dx^{n}}[(x-\alpha)^{m}e^{-\beta x}]=e^{-\beta x}\sum_{i=0}^{n}(-\beta)^{n-i}\nonumber\\
	\frac{m!}{(m-i)!}\frac{n!}{(n-i)!i!}(x-\alpha)^{m-i},n\le m
\end{align}
which can be proved by the mathematical induction.
From Eq.~(\ref{S1_10}) to Eq.~(\ref{S1_11}), 
$m_1-m_4$ is substituted by $K$ and $m_4-u$ is substituted by $s$. 
In Eq.~(\ref{S1_12}), the associated Laguerre polynomial $\mathcal{L}_{m_1}^{(K)}$ can be 
expressed in the Rodrigues representation
\begin{align}\label{Laguerre}
	L_{n}^{(k)}(x)=\sum_{s=0}^{n}\frac{(-1)^{s}(n+k)!x^{s}}{(n-s)!(s+k)!s!}.
\end{align}
When $m_1>m_4$, the derivation is similar except that the derivative 
with respect to $m_1$ is evaluated earlier than that with respect to $m_4$.
We arrive at 	
\begin{align}\label{S3S12}
	S_1({\bm q})=&\sqrt{\frac{m_4!}{m_1!}}
	\sum_{\Delta n}\delta_{q_x,\frac{2\pi}{L_x}[j_1-j_4+\Delta nm_L]}\nonumber\\
	&e^{-q^2l_B^2/4-iq_{z}(Z_{j_{1}}+\Delta nL_z+Z_{j_{4}})/2}\nonumber\\
	&[\frac{(q_{x}+iq_{z})l_B}{\sqrt{2}}]^{K}\mathcal{L}_{m_1}^{(K)}(\frac{q^2l_B^2}{2}).
\end{align}
Combining Eq.~(\ref{S1_12}) and Eq.~(\ref{S3S12}), we get
\begin{align}\label{S3S1Final}
		S_1({\bm q})=\sqrt{\frac{\min{(m_1,m_4)}!}{\max{(m_1,m_4)}!}} 		
		\sum_{\Delta n}\delta_{q_x,\frac{2\pi}{L_x}[j_1-j_4+\Delta nm_L]}&\nonumber\\
		e^{-q^2l_B^2/4-iq_{z}(Z_{j_{1}}+\Delta nL_z+Z_{j_{4}})/2}&\nonumber\\
		[\frac{\mbox{sgn}(m_{1}-m_{4})q_{x}+iq_{z}}{\sqrt{2}}l_B]^{|m_{1}-m_{4}|}L_{\min(m_{1},m_{4})}^{(|m_{1}-m_{4}|)}(\frac{q^{2}l_B^{2}}{2}).&
\end{align}
In the same way, we obtain
\begin{align}\label{S3S2Final}
		S_2({\bm q})=\sqrt{\frac{\min{(m_2,m_3)}!}{\max{(m_2,m_3)}!}} 		
		\sum_{\Delta n^\prime}\delta_{q_x,\frac{2\pi}{L_x}[j_3-j_2+\Delta n^\prime m_L]}\nonumber\\
		e^{-q^2l_B^2/4+iq_{z}(Z_{j_{2}}+\Delta n^\prime L_z+Z_{j_{3}})/2}\nonumber\\
		[\frac{-\mbox{sgn}(m_{2}-m_{3})q_{x}-iq_{z}}{\sqrt{2}}l_B]^{|m_{2}-m_{3}|}L_{\min(m_{2},m_{3})}^{(|m_{2}-m_{3}|)}(\frac{q^{2}l_B^{2}}{2}).
\end{align}
Combining Eq.~(\ref{S3S1Final}) and Eq.~(\ref{S3S2Final}), we have
\begin{widetext}
\begin{align}
	S_1({\bm q})S_2({\bm q})\delta_{q_x,\frac{2\pi}{L_x}t}
	\delta_{q_z,\frac{2\pi}{L_z}s}&= G_1({\bm q})G_2({\bm q}) \sum_{\Delta n\Delta n^\prime}
	\delta_{q_x,\frac{2\pi}{L_x}[j_1-j_4+\Delta nm_L]}
	\delta_{q_x,\frac{2\pi}{L_x}[j_3-j_2+\Delta n^\prime m_L]}
	\notag 
	\\
	&e^{iq_{z}(Z_{j_{2}}+Z_{j_{3}}-Z_{j_{1}}-Z_{j_{4}}+\Delta n^\prime L_z+\Delta nL_z)/2}
	\label{S3SFinal_1}\\
	&= G_1({\bm q})G_2({\bm q})\sum_{\Delta n\Delta n^\prime}\delta_{t,j_1-j_4+\Delta nm_L}
	\delta_{j_1-j_4+\Delta nm_L,j_3-j_2+\Delta n^\prime m_L}\notag\\
	&e^{i\frac{\pi s(j_2+j_3-j_1-j_4)}{m_L}+is\pi(\Delta n+\Delta n^\prime)}\label{S3SFinal_2}\\
	&=G_1({\bm q})G_2({\bm q})\sum_{\Delta n\Delta n^\prime}\delta_{t,j_1-j_4+\Delta nm_L}
	\delta_{j_2-j_4,j_3-j_1+(\Delta n^\prime-\Delta n) m_L}	\notag\\
	&e^{i\frac{2\pi s(j_3-j_1)}{m_L}+is\pi(\Delta n+\Delta n^\prime+\Delta n^\prime-\Delta n)}\label{S3SFinal_3}\\
	&=G_1({\bm q})G_2({\bm q})\delta_{t,j_1-j_4}^\prime
	\delta_{j_1+j_2,j_3+j_4}^\prime
	e^{i\frac{2\pi s(j_3-j_1)}{m_L}},\label{S3SFinal_4}
\end{align}
\end{widetext}
where the prime on the $\delta$ function represents that the equality is defined module $m_L$
and 
\begin{align}\label{S3G1}
	&G_1({\bm q})=\sqrt{\frac{\min{(m_1,m_4)}!}{\max{(m_1,m_4)}!}} e^{-q^2l_B^2/4}&\nonumber\\
	&[\frac{\mbox{sgn}(m_{1}-m_{4})q_{x}+iq_{z}}{\sqrt{2}}l_B]^{|m_{1}-m_{4}|}L_{\min(m_{1},m_{4})}^{(|m_{1}-m_{4}|)}(\frac{q^{2}l_B^{2}}{2}),	&	
\end{align}
\begin{align}\label{S3G2}
	&G_2({\bm q})=\sqrt{\frac{\min{(m_2,m_3)}!}{\max{(m_2,m_3)}!}} e^{-q^2l_B^2/4}&\nonumber\\
&	[\frac{-\mbox{sgn}(m_{2}-m_{3})q_{x}-iq_{z}}{\sqrt{2}}l_B]^{|m_{2}-m_{3}|}L_{\min(m_{2},m_{3})}^{(|m_{2}-m_{3}|)}(\frac{q^{2}l_B^{2}}{2}).	&	
\end{align} 
In summary, 
\begin{widetext}
\begin{equation}\label{S3AFinal}
	A_{j_1j_2j_3j_4}^{m_1h_1m_2h_2m_3h_3m_4h_4}=\frac{1}{2L_xL_z}
	\sum_{q_x,q_z,s,t}^\prime \frac{2\pi}{q}F({\bm q})G_1({\bm q})G_2(\bm q)
	e^{i\frac{2\pi s(j_3-j_1)}{m_L}}
	\delta_{q_x,\frac{2\pi}{L_x}t}
	\delta_{q_z,\frac{2\pi}{L_z}s}
	\delta_{t,j_1-j_4}^\prime
	\delta_{j_1+j_2,j_3+j_4}^\prime.
\end{equation}
\end{widetext}

 \section{Appendix D: THE REAL-SPACE DENSITY distribution at a fixed $y$}
In the main text, we show that the real-space density distribution of the calculated FQH liquid 
averaged over $y$ is nearly uniform in the $(x, z)$ plane.
In this Appendix, we will provide the density distribution at any fixed $y$ to illustrate its near uniformity.

\begin{figure*}[t]
	\includegraphics[width=6.5in]{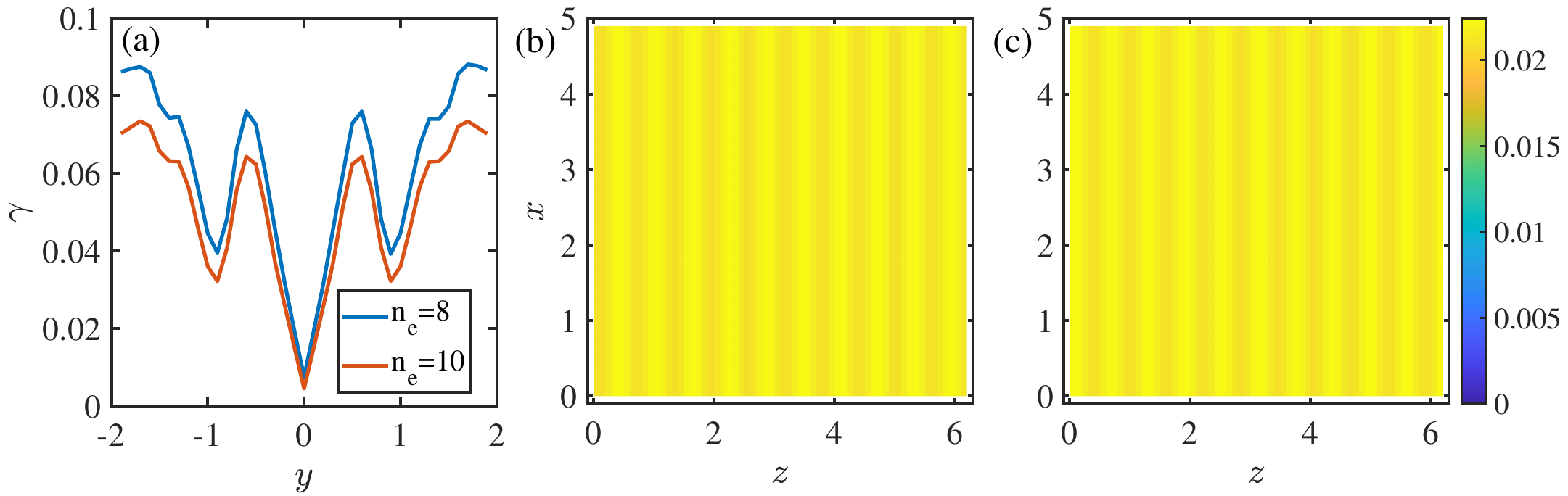}
	\caption{
		(a) The fluctuation of the real-space density for 8 (blue line) and 10 (red line) electrons.
		Here we show the result of the ground state with $K_x=20$ for $n_e=8$ and $K_x=25$ for $n_e=10$, and the results of the other ground states are the same. 
		The real-space density distribution in the $(x,z)$ plane at (b) $y=-1.7$ and (c) $y=1.7$ for the ground state with $K_x=25$ for $n_e=10$.
		The aspect ratio is $r_a=1$ for $n_e=8$ and $r_a=0.8$ for $n_e=10$, at which
		the gap between the triply degenerate ground states and excited states is maximum.}
	\label{FigS3}
\end{figure*}

We calculate the fluctuation of the real-space density distribution in the $(x,z)$ plane at a fixed value of $y$,
\begin{equation}\label{S4fluc}
	\gamma(y)=\frac{ \rho_{\max}(y)-\rho_{\min}(y)}{\overline{\rho}(y)},
\end{equation}
where $\rho_{\max}(y),\rho_{\min}(y)$ and $\overline{\rho}(y)$ represent the maximum, minimum
and average value of the density $\rho({\bm r})$ over the $(x,z)$ plane at a specific $y$, respectively.
The result is shown in Fig.~\ref{FigS3}
where we see that the fluctuation is small, within 9\% at the maximum point for $n_e=8$ electrons and within 8\%
for $n_e=10$ electrons.
The fluctuation for $n_e=10$ is smaller than that for $n_e=8$, suggesting that the finite fluctuation 
is due to the finite-size effect. 
In addition, we plot the density in the $(x,z)$ plane near the bottom and top surfaces for $n_e=10$ electrons
in Fig.~\ref{FigS3}(b) and (c)
and find that the density distributions exhibit a slight stripe pattern.
Note that the single-particle states in our calculations are also localized along $z$ and
extended along $x$ due to our gauge choice, supporting that the stripe pattern in 
the density distribution of the many-body ground state is attributed to our basis choice for the finite-size system.

\section{Appendix E: THE MANY-BODY CHERN NUMBER}
In the main text, we calculate the many-body Chern number by imposing twisted boundary 
phases ${\bm \theta}=(\theta_x,\theta_z)$~\cite{niu1985PRB}, 
which transforms $k_x$ to $k_x-\theta_x/L_x$ and $k_z$ to $k_z-\theta_z/L_z$ in the single-particle
Hamiltonian (\ref{S1Ham2}).
In terms of $\hat{b}^\dagger$ and $\hat{b}$ in Eq.~(\ref{S1Ope1}) and Eq.~(\ref{S1Ope2}),
we obtain 
\begin{equation}\label{S4Ham1}
	H_B=\left(\begin{array}{cc}
		H_{11}
		& A[\frac{\hat{b}^\dagger +\hat{b}}{\sqrt{2}l_B}-\partial_y-\frac{\theta_x}{L_x}]\\
		A[\frac{\hat{b}^\dagger +\hat{b}}{\sqrt{2}l_B}+\partial_y-\frac{\theta_x}{L_x}]& 	H_{22}
	\end{array}\right).
\end{equation}
where
\begin{align}\label{S4H11}
	H_{11}&=	\frac{2(D_2-M)}{l_B^2}[\hat{b}^\dagger \hat{b}+\frac{1}{2}-\frac{l_B\theta_x}{\sqrt{2}L_x}(\hat{b}+\hat{b}^\dagger)\nonumber\\
&+i\frac{l_B\theta_z}{\sqrt{2}L_z}(\hat{b}-\hat{b}^\dagger)+(\frac{\theta_x}{L_x})^2+(\frac{\theta_z}{L_z})^2]\nonumber\\
&	-(D_{1}-M)\partial ^{2}_y+Mk_{w}^{2} ,
\end{align}	
\begin{align}\label{S4H22}
	H_{22}&=	\frac{2(D_2+M)}{l_B^2}[\hat{b}^\dagger \hat{b}+\frac{1}{2}-\frac{l_B\theta_x}{\sqrt{2}L_x}(\hat{b}+\hat{b}^\dagger)\nonumber\\
&+i\frac{l_B\theta_z}{\sqrt{2}L_z}(\hat{b}-\hat{b}^\dagger)+(\frac{\theta_x}{L_x})^2+(\frac{\theta_z}{L_z})^2]\nonumber\\
&-(D_{1}+M)\partial ^{2}_y-Mk_{w}^{2} .
\end{align}	
The single-particle eigenstate wave function is  $[\phi^u_{0,j}({\bm \theta}),\phi^u_{1,j}({\bm \theta})]^T$ where
\begin{equation}\label{S4WF}
	\phi^u_{\sigma,j}(x,y,z,{\bm \theta})=\sum_{m,h}c_{m,h,\sigma}^u({\bm \theta})\psi_{m,h,j}(x,y,z)
\end{equation}	
with the basis $\psi_{m,h,j}(x,y,z)$ the same as (\ref{S1Basis4}).
By projecting the Coulomb interaction onto one Landau level under the boundary phase 
${\bm \theta}$ and diagonalizing the many-body Hamiltonian, 
we have the many-body ground state
\begin{equation}\label{S4State}
	|\Psi_M({\bm \theta})\rangle=\sum_{\{j\}}O_{\{j\}}({\bm \theta})\{\hat{a}_j^\dagger({\bm \theta})\}
	|0\rangle,
\end{equation}
where $\{\hat{a}_j^\dagger({\bm \theta})\}=\hat{a}_{j_1}^\dagger({\bm \theta})
\hat{a}_{j_2}^\dagger({\bm \theta})... \hat{a}_{j_{n_e}}^\dagger({\bm \theta})$
is a sequence of  creation operators creating the Fock basis with $\hat{a}_j^\dagger({\bm \theta})$	corresponding to the 
single-particle state $\phi^u_{\sigma,j}(x,y,z,{\bm \theta})$ on the specific Landau level under
the boundary phase ${\bm \theta}$. 
$O_{\{j\}}({\bm \theta})$ is the coefficient for each Fock basis with $\{j\}=j_1j_2...j_{n_e}$
and $n_e$ being the number of electrons.
To evaluate the many-body Chern number, we use the method proposed in~\cite{fukui2005JPSJ} for the 
discrete Brillouin zone:
	\begin{align}\label{S4Chern}
		C=\frac{1}{2\pi i}\sum_{\bm {\theta_0}}\ln \left[ \right.
		&\langle \Psi_M(\bm{ \theta_0})|\Psi_M(\bm{ \theta_0}+\hat{\bm{ \theta}}_{\bm x})\rangle\nonumber \\
	&	\langle 
		\Psi_M(\bm{ \theta_0}+\hat{\bm{ \theta}}_{\bm x})|\Psi_M(\bm{ \theta_0}+\hat{\bm{ \theta}}_{\bm x}+\hat{\bm{ \theta}}_{\bm y})\rangle\nonumber \\
	&	\langle \Psi_M(\bm{ \theta_0}+\hat{\bm{ \theta}}_{\bm x}+\hat{\bm{ \theta}}_{\bm y})|\Psi_M(\bm{ \theta_0}+\hat{\bm{ \theta}}_{\bm y})\rangle\nonumber \\
	&	\left.\langle \Psi_M(\bm{ \theta_0}+\hat{\bm{ \theta}}_{\bm y})|\Psi_M(\bm{ \theta_0})\rangle \right]
	\end{align}
where $\hat{\bm{ \theta}}_{\bm x}=(2\pi/N_x,0),\hat{\bm{ \theta}}_{\bm z}=(0,2\pi/N_z)$ with 
$N_x,N_z$ taking integer values and $\bm{ \theta_0}=(2\pi s/N_x,2\pi t/N_z)$ with
$s=0,1,2,...,N_x-1$ and $t=0,1,2,...,N_z-1$. Using Eq.~(\ref{S4WF}) and Eq.~(\ref{S4State}), we get
\begin{align}\label{S4Prod}
	\langle \Psi_M(\bm{ \theta_1}|\Psi_M(\bm{ \theta_2})\rangle=&
	\sum_{\{j\}}O^*_{\{j\}}(\bm {\theta_1})O_{\{j\}}(\bm {\theta_2})\nonumber\\
&	[\sum_{m,h,\sigma}c_{m,h,\sigma}^{*u}(\bm{\theta_1})c_{m,h,\sigma}^{u}(\bm{\theta_2})]^{n_{e}}.
\end{align}
For the calculation in the main text, we take $N_x=N_z=20$.

\section{Appendix F: THE PARTICLE-ENTANGLEMENT SPECTRUM}
In this Appendix, we will review the method for calculating the particle-entanglement spectrum~\cite{sterdyniak2011PRL,regnault2011PRX} in detail,
especially on how to decompose the Fock basis to partition the electrons into two parts.

To calculate the particle-entanglement spectrum of a many-body state $|\Psi\rangle$ with $N$ orbitals and $n_e$ electrons, 
we divide the $n_e$ electrons into $A$ and $B$ part with $n_a$ and $n_b$ electrons, respectively,
where $n_a+n_b=n_e$.
Tracing out the electrons in part $B$, we obtain the reduced density matrix 
\begin{equation}\label{S5rho}
	\rho_A=\mbox{Tr}_B(|\Psi\rangle\langle \Psi|),
\end{equation}
the eigenvalues of which are just the entanglement spectrum.
Taking the partial trace in terms of two parts of particles is not as straightforward as two blocks of 
orbitals.
The key point is to decompose each Fock basis with $N$ orbitals and $n_e$ electrons into the tensor
product of two other Fock basis, both with $N$ orbitals while one with $n_a$ electrons and another with
$n_b$ electrons.
We will explain in detail in the following.

We use a sequence $\{j\}=[j_1j_2...j_{n_e}]$ to represent a Fock basis with $N$ orbitals and $n_e$ electrons, for which the $j_i$th orbital is occupied and we place the $j_i$ in an ascending order.
Obviously, $j_i \in \{1,2,...,N\}$ and all the $j_i$ in the same sequence are different from each other.
For example, for $N=6$ and $n_e=3$, the sequence $\{j\}=[256]$ represents the Fock basis $|010011\rangle$.
For the purpose of dividing the electrons into two parts, $\{j\}$ is decomposed as
\begin{equation}\label{S5decompose}
	\{j\}=\frac{1}{\sqrt{N_p}}\sum_{\{k\}}(-1)^t\{k\}\{l\},
\end{equation}
where 
$\{k\}$ and $\{l\}$ are also sequences representing Fock basis with $\{k\}=[k_1k_2...k_{n_a}]$,
$\{l\}=[l_1l_2...l_{n_b}]$ and $k_i,l_i \in \{1,2,...,N\}$, and ${N_p}=\left(\begin{array}{c}
	n_e\\
	n_a
\end{array}\right)$ denotes the number of the divided parts.
The combination of $\{k\}$	and the corresponding $\{l\}$, i.e. $\{k\}\cup\{l\}=[k_1k_2...k_{n_a}l_1l_2...l_{n_b}]$ must recover $\{j\}$ up to a permutation.
If an odd permutation is needed to recover $\{j\}$, $t$ is odd and otherwise, $t$ is even.
Let us explain the method by an explicit example.
If we decompose the basis $\{j\}=[256]$ with $N=6$ orbitals into two parts with $n_a=1$ and $n_b=2$ electrons, 
respectively,
\begin{equation}\label{S5decomeg}
	[256]=\frac{1}{\sqrt{3}}([2][56]-[5][26]+[6][25]),
\end{equation}
representing 
\begin{align}\label{S5decomFock}
	|010011\rangle=&\frac{1}{\sqrt{3}} ( |010000\rangle|000011\rangle-|000010\rangle|010001\rangle\nonumber\\
	&+|000001\rangle|010010\rangle).
\end{align}
It is easy to check the equality in Eq.~(\ref{S5decomFock}) by writing both sides in the form of 
first-quantized wave functions.
As the many-body state $|\Psi\rangle$ is a superposition of the Fock basis with $N$ orbitals and $n_e$ electrons,
it can be written in the form $|\Psi\rangle=\sum_{\{j\}\{k\}}c_{\{j\}}(-1)^t\{k\}\{l\}$ where
$c_{\{j\}}$ is the coefficient before the basis $\{j\}$ after normalization.
By tracing out the $B$ part with $n_b$ electrons, i.e. the $\{l\}$, we get the reduced density matrix $\rho_A$ and 
then the particle-entanglement spectrum by the exact diagonalization.

\bibliography{FQHE.bib}

\end{document}